\documentclass[]{spie}  

\usepackage{amsmath,amsfonts,amssymb}
\usepackage{graphicx}
\usepackage[colorlinks=true, allcolors=blue]{hyperref}

\title{Noise modeling and analysis of an IMU-based attitude sensor: improvement of performance by filtering and sensor fusion}

\author[a]{K.~Nirmal}
\author[a]{A.~G.~Sreejith}
\author[a]{Joice Mathew}
\author[a]{Mayuresh Sarpotdar}
\author[a]{Ambily Suresh}
\author[a]{Ajin Prakash}
\author[a]{Margarita Safonova}
\author[a]{Jayant Murthy}
\affil[a]{Indian Institute of Astrophysics, Bangalore, India}

\authorinfo{Further author information: (Send correspondence to K.~Nirmal)\\ E-mail: nirmalk@iiap.res.in}

\begin{document} 
\maketitle

\begin{abstract}
We describe the characterization and removal of noise present in the Inertial Measurement Unit (IMU) MPU-6050. This IMU was initially used in an attitude sensor (AS) developed in-house, and subsequently implemented in a pointing and stabilization platform developed for small balloon-borne astronomical payloads. We found that the performance of the IMU degrades with time due to the accumulation of different errors. Using the Allan variance analysis method, we identified the different components of noise present in the IMU and verified the results using a power spectral density analysis (PSD). We tried to remove the high-frequency noise using smoothing filters, such as moving average filter and Savitzky-Golay filter. Although we did manage to filter some of the high-frequency noise, the performance of these filters was not satisfactory for our application. We found the distribution of the random noise present in the IMU using a probability density analysis, and identified the noise to be white Gaussian in nature which we successfully removed by a Kalman filter in real time. 
\end{abstract}

\keywords{Balloon experiment, Attitude sensor, Pointing system, MEMS sensors.}

\section{INTRODUCTION}
\label{sec:intro}  
\begin{figure}[ht]
\centering
\includegraphics[scale=0.4]{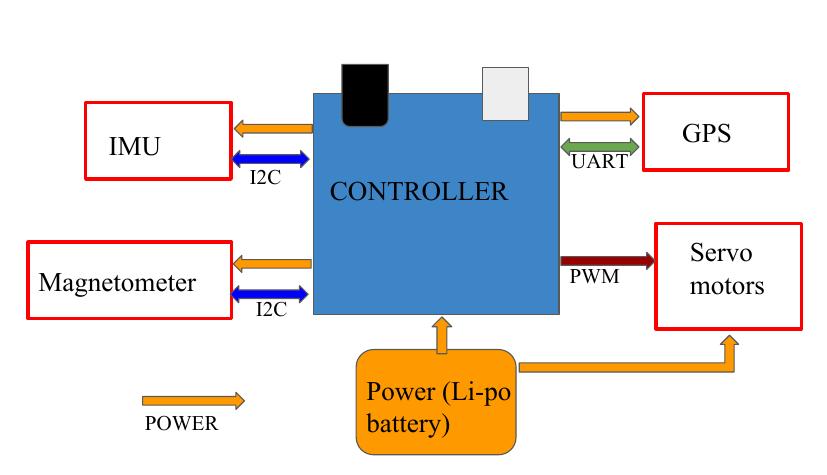}
\includegraphics[scale=0.3]{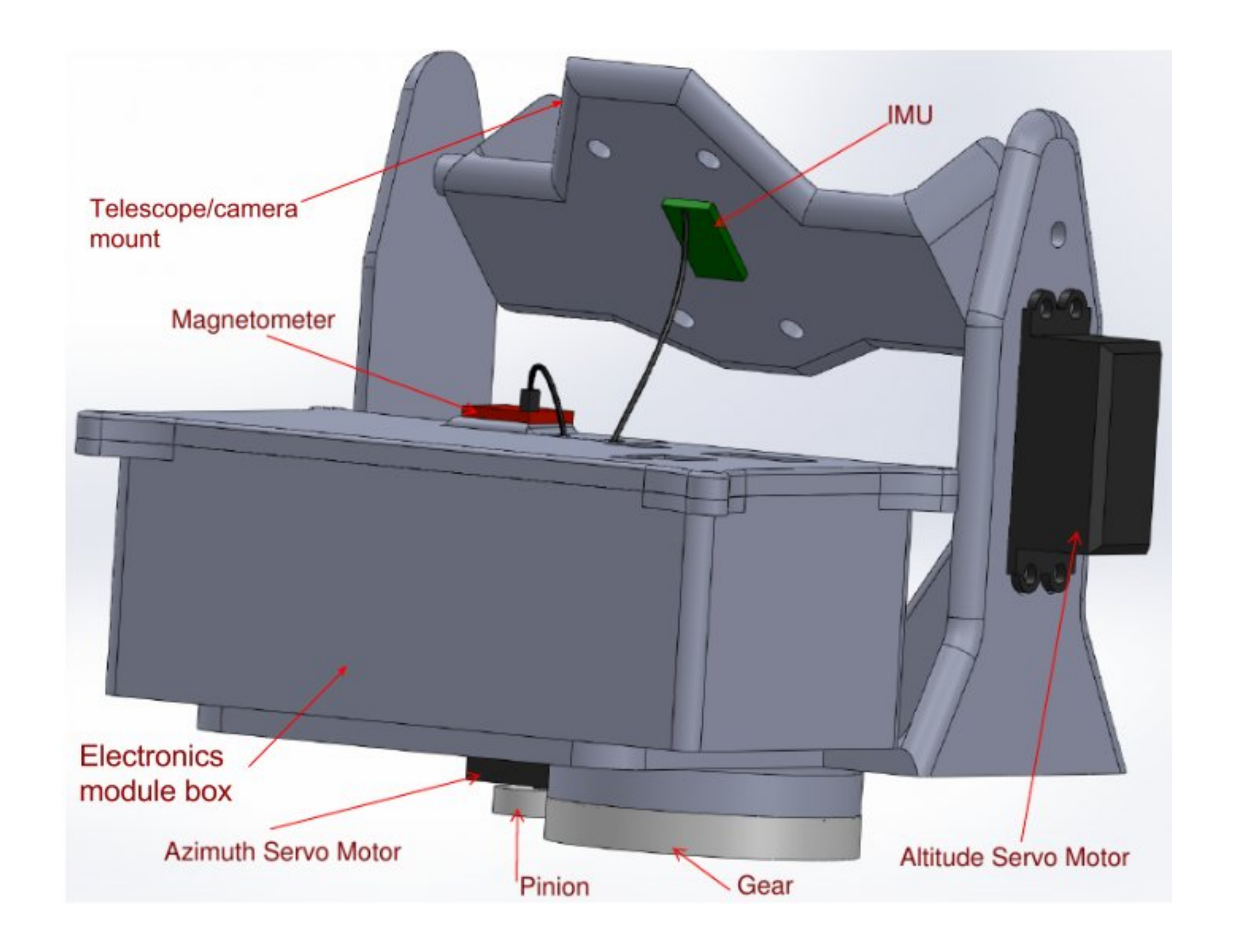}
\caption{{\it Left}: A block diagram of the pointing system. {\it Right}: SOLIDWORKS rendering of the pointing system.}
\label{fig:ps}
\end{figure}

Balloon experiments are an economically feasible method of conducting experiments or observations in astronomy that are not possible from the ground. The payload in these experiments may include a telescope, a detector, and a pointing system/stabilization platform (gondola) [\citenum{Crill08}]. In balloon-borne astronomical payloads, it is essential to determine the attitude of the payload to point the detector/telescope to the desired direction, for which several attitude determination techniques are usually used: star sensors, sun sensors, IMUs, gyroscopes and/or magnetometers, whose weight might amount to several kilograms [\citenum{Pascale}]. Since our balloon experiments are constrained by the weight of a payload (lightweight category), we have developed a Raspberry Pi (RPi) based attitude sensor (AS) [\citenum{Sreejith14}]. The weight of this AS is only several grams making it ideal to use in lightweight balloon experiments. We  modified this AS into the pointing and stabilization system (Fig.~\ref{fig:ps}) by replacing the RPi with the Arduino controller (more suitable for pointing correction than the RPi) and adding the  servomotors for attitude control. The duration of our balloon flight spans 3 to 4 hours, and the IMU-based AS is used for the entire length of the flight to measure the payload orientation. However, the reliability of the IMU response of degrades with time due to accumulation of errors. Therefore, the errors in the IMU should be measured and accurately calibrated before the flight. Some errors in the IMU response, such as static biases, sensor misalignment, and scale factor errors, are comparatively easy to measure and calibrate. In contrast, stochastic errors, such as measurement noise, drifting biases, and turn-on to turn-on bias variation, are based around random processes and thus difficult to model.

Here, we discuss modelling and correcting the IMU measurement noise. We have used methods of Allan variance (AV) [\citenum{Allan}] and Power Spectral Density (PSD) analysis to identify and verify the nature of errors in the IMU response, and modelled the errors using the probability density function (PDF) analysis.

We implemented a low-pass moving average filter and a Savitzky-Golay smoothing filter in the controller to filter the noise in real-time (in the working pointing system), but that did not give satisfactory results. There was a lag between the filtered data and the actual response of the IMU. The application of the real-time Kalman filter in our controller, on the other hand, was successful. 

In Section~\ref{sec:IMU} we describe the IMU and determination of angular displacement using the Arduino controller. Different techniques to characterize the measurement noise in the IMU such as AV analysis, PSD analysis, and PDF analysis are described in Sections~\ref{sec:av} to \ref{sec:pdf}. In Section~\ref{sec:filter} we discuss different filtering techniques implemented, their performances and limitations, and conclude with the IMU calibration in Section~\ref{sec:calib}.

\section{IMU MPU-6050 AND ESTIMATION OF ANGULAR DISPLACEMENT}
\label{sec:IMU}

IMU MPU-6050 comprises a combination of a 3-axis accelerometer and a 3-axis gyroscope, and a digital motion processor capable of motion fusion. Here, the data from different sensors is either combined to derive the attitude, or can be used separately. In principle, a better accuracy is achieved by fusing data from the  individual sensors [\citenum{Emilson}].
The accelerometer and gyroscope data are digitized by two 16-bit ADC in the IMU, and sent to the Arduino through an I2C port at a rate of 400 KHz. The I2C port is a multi-master serial single-ended computer bus, through which low-speed peripherals attached to the controller. 

A MEMS gyroscope consists of vibrating solid state resonators,  which maintain the plane of vibration even if the gyroscope is tilted or rotated. This Coriolis vibratory gyroscope (CVG) is common in electronic gadgets such as tablets and mobile phones. A voltage, proportional to the angular velocity of the IMU, is generated in the gyroscope and is digitized by a 16-bit ADC, whose full scale reading corresponds to 3.3 V. Thus, the analog voltage generated for an angular velocity of $1^{\circ}$/sec is 3.3  mV/($^{\circ}$/sec) 
[\citenum{mpu_datasheet}], corresponding to an ADC value of $\frac{3.3\,mV}{3.3\,V} \times (2^{16}-1 ) = 65.535$. The gyroscope generates a bias voltage which is measured when the IMU is stationary, and this bias is subtracted from the ADC value to get the actual response. The angular velocity rate in degrees per second is calculated as
\begin{eqnarray}
\mbox{Angular velocity rate} (\omega) &=& \frac{(V_{\rm ADC} - V_{\rm bias})}
{\textit{sensitivity}}\,  (^{\circ}/\rm sec) \nonumber\\
& = &\frac{(V_{\rm ADC}- V_{\rm bias})}{65.535}\, (^{\circ}/\rm sec)\,.
\label{eq:omega}
\end{eqnarray}
and the angular displacement ($\theta$) is calculated  by multiplying the rate by the time period $\Delta t$ ($\theta = \omega \times \Delta t$).

The accelerometer measures the acceleration (g) in the $X$, $Y$ and $Z$ axes and generates a voltage proportional to the acceleration. The bias voltage ($1.5\,V$), inherent in the  accelerometer ADC output, is subtracted from the accelerometer output to get voltage corresponding to the acceleration,
\begin{eqnarray}
V_{acc} = V_{\rm ADC} - V_{\rm bias}\,,
\end{eqnarray}
where $V_{acc}$, $V_{\rm ADC}$, $V_{\rm bias}$ are the voltages corresponding to actual acceleration reading, ADC output and bias, respectively.

The angular displacement in degrees is found from acceleration values using the equation
\begin{eqnarray}
\theta = \arctan \left(\frac{{V_{accy}}}
{{V_{accz}}}\right) + \pi\,,
\label{eq:alt}
\end{eqnarray}
where $V_{accy}, V_{accz}$ are the bias-subtracted accelerometer ADC outputs corresponding to acceleration in $Y$ and $Z$-axis, respectively.

\section{ANALYSIS OF NOISE IN THE IMU}

The IMU errors can be divided into two categories: deterministic and stochastic. 
Deterministic noise arises due to such factors as misalignment of the sensor chip during fabrication, static bias or scale factor errors. The latter is the gain value of a sensor which relates acceleration or rotation to the produced voltage. Deterministic errors remain either constant or vary over a period in such a way that they can be determined by a mathematical model.

However, there are other errors in IMU response that are completely random in nature, known as stochastic errors. It is only possible to represent these errors as random variables having some probabilistic distribution. Different types of stochastic errors in the IMU are the measurement noise, drifting biases, and turn-on to turn-on bias variance. 

\subsection{MEASUREMENT NOISE}

These are the errors that arise while measuring the output of the IMU. They include quantization noise, angle/velocity random noise, flicker noise, angular rate/acceleration random walk and ramp noise. Two properties of these random errors should be understood to model them correctly --  the colour and the probability density function (PDF). According to the Power Spectral Density (PSD) method, noise can be classified as white noise (where PSD is flat at all frequencies), pink noise (where PSD $\propto 1/f$), and brown noise/red noise (where PSD $\propto 1/f^2$). Several techniques, such as autocorrelation function method, power spectral density (PSD) analysis approach and Allan Variance (AV), are used to characterize these errors [\citenum{Zaho}]. We have used the AV analysis to identify the measurement noise in the IMU signal, and this was further verified by the PSD analysis.

\subsection{AV ANALYSIS}
\label{sec:av}

\begin{figure}[ht]
\centering
\includegraphics[scale=0.08]{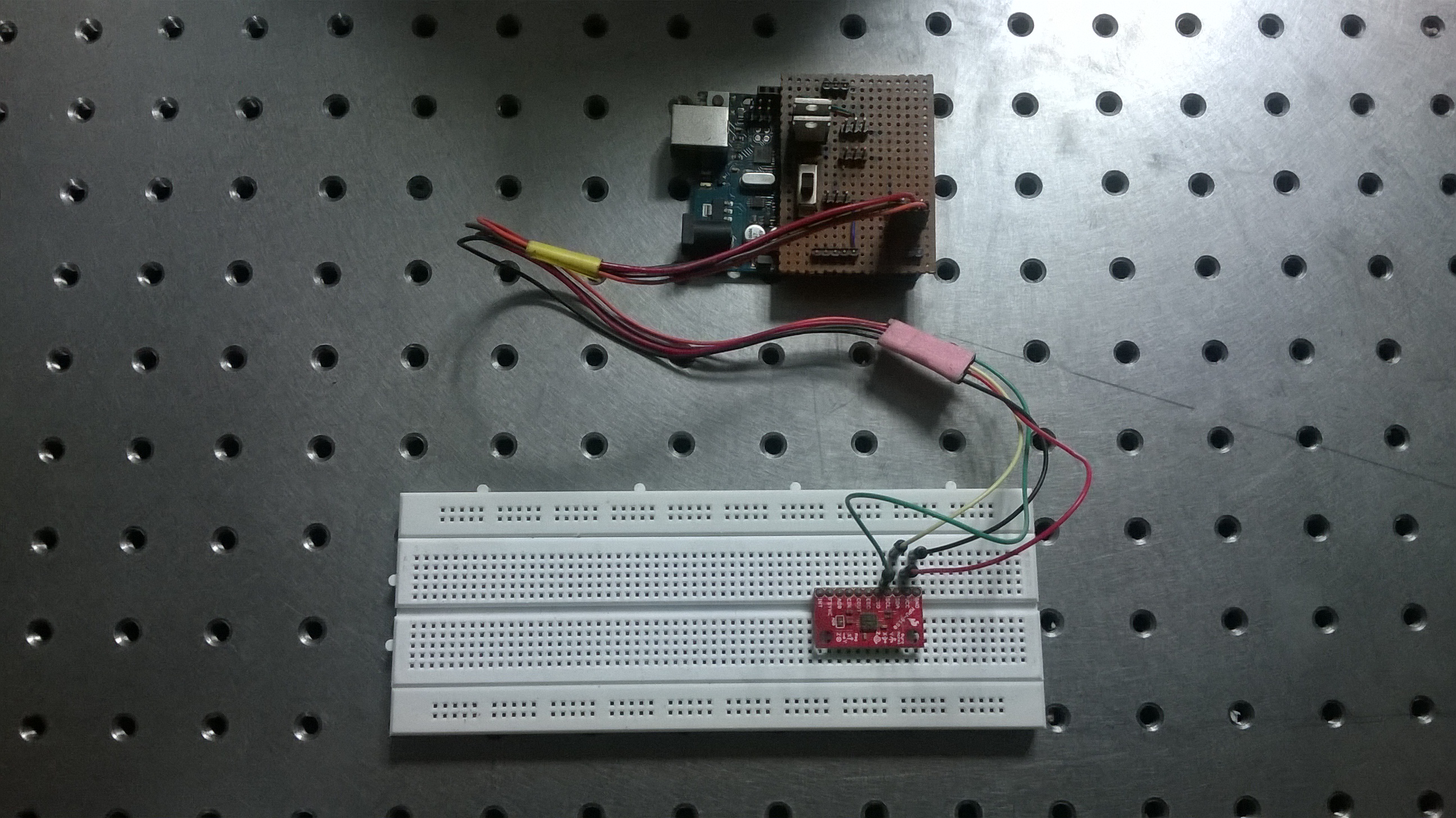}
\caption{The experimental setup take the accelerometer and gyroscope readings. The IMU is kept stationary by keeping it on the optical table.}
\label{fig:expt_setup}
\end{figure}

The AV technique is used to identify
and quantify various noise terms that exist in
the data and to determine the colour or the combination of colours that can be used to explain the noise process. The data from the accelerometer and gyroscope are collected from the stationary IMU through an Arduino controller I2C port, and saved to a computer, connected to the Arduino through a serial port. The IMU is kept on the optical table (Fig.~\ref{fig:expt_setup}) during the data collection to ensure the stationary condition.  

\begin{figure}[h!]
\centering
\includegraphics[width=5in]{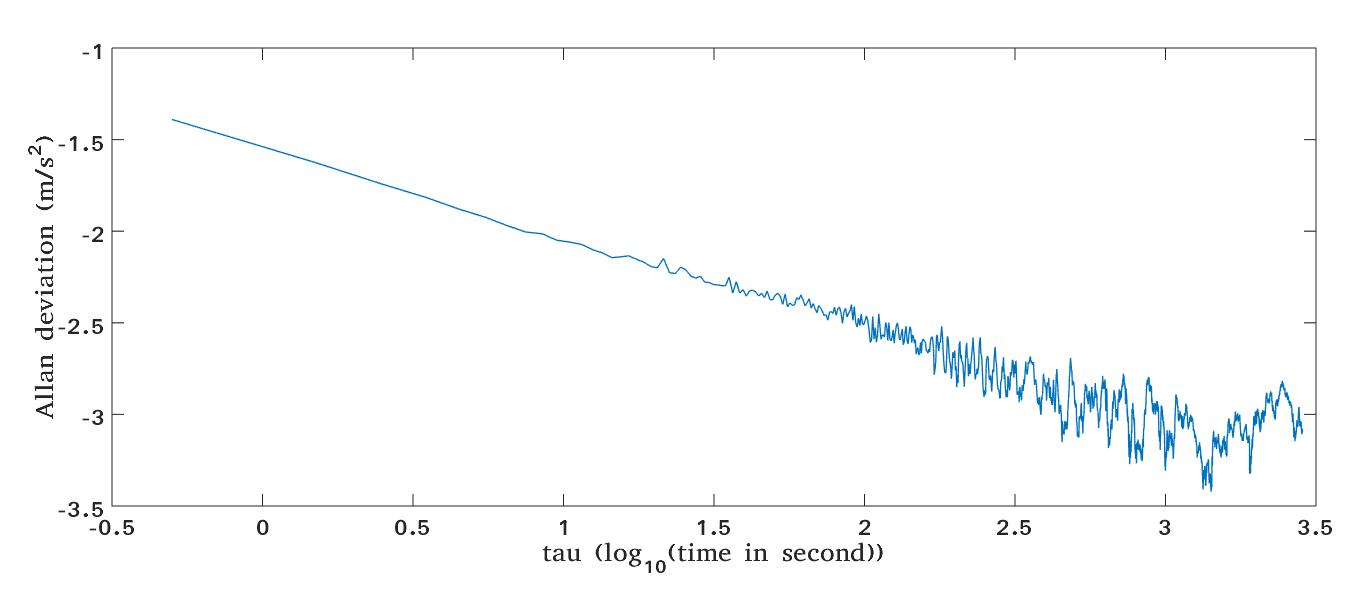}
\caption{Allan deviation function of accelerometer's $X$-axis data. The slope was measured by fitting a line to this curve. The axes are in log scale.}
\label{fig:av_analysis}
\end{figure}

An $N$-point equally spaced time series data collected from the IMU was subject to the Allan Variance analysis as follows. The data were divided into $m$ equal segments each containing $n$ elements. The time spanned by each segment is $\tau = n \times \Delta t $. The time average of each segment is represented as $s_{11}(\tau),\, s_{12}(\tau), \, s_{13}(\tau),\dots s_{1m}(\tau)$. These averages were used to calculate the Allan variance,
\begin{equation}
\sigma^{2}(\tau)  = \frac{1}{(2s -1)}\sum_{i=1}^{s-1} (s_{i+1}(\tau)-s_{i}(\tau))^{2} \,.
\label{eq:AV_eq}
\end{equation}
We repeated this process for different values of $n$ ranging from $n = 1$ to at least $n=N/10$. Once Allan variance is calculated for different $\tau$, Allan deviation (AD) is computed by taking the square root of the Allan variance. The time $\tau$ and the Allan deviation are plotted on the log-log scale graph (Fig.~\ref{fig:av_analysis}), known as the Allan deviation function (ADF). Different types of noise produce different Allan deviation functions. The colour of the noise can be found from the Allan deviation function slope [\citenum{Zaho}]. In Table~\ref{tab:noise_para} we display the characteristics of the different types of the noise. 

In our analysis, the curves obtained from the Allan variance show similar characteristics. When $\tau$ is small, the  Allan variance function is decreasing, which is the characteristic of a measurement noise. At larger $\tau$, in some cases the slope changes from $-1/2$ to $+1/2$, indicating the presence of a drifting bias noise. Drifting biases are prevalent when the data samples are large (in hours). We have measured the slope of the Allan deviation function (given in the second column in Table~\ref{tab:AV_AN}) and determined that the IMU measurement noise is the white noise. 

\begin{table}[h!]
\caption{Relation between Allan deviation function and noise coefficients} 
\label{tab:noise_para}  
\begin{center}  
\begin{tabular}{|l|c|c|c|}
\hline
\rule[-1ex]{0pt}{3.5ex}  Noise type & \begin{tabular}[x]{@{}c@{}}AD $\left( \sigma(\tau)\right)$ \end{tabular}  & \begin{tabular}[x]{@{}c@{}}Slope of\\ ADF \end{tabular}   & \begin{tabular}[x]{@{}c@{}}Noise\\coefficient $Q_{z}$\end{tabular} \\
\hline
\rule[-1ex]{0pt}{3.5ex}

\begin{tabular}[x]{@{}c@{} c@{}} White noise\\ (Angle/velocity \\ random walk)\end{tabular} &  $ \frac{Q_{z}}{\sqrt{\tau}}$   & $-1/2$ &  $\sigma\left(1\right)$\\
\hline

\rule[-1ex]{0pt}{3.5ex}  
\begin{tabular}[x]{@{}c@{} c@{}} Quantization \\ noise \end{tabular} &  $ \frac{\sqrt{3} \times Q_{z}}{\tau}$ & $-1$ & $ \sigma\left(\sqrt{3}\right)$\\
\hline
\rule[-1ex]{0pt}{3.5ex}  
\begin{tabular}[x]{@{}c@{}}Drifting bias\\ (Rate random walk) \end{tabular} & $\frac{Q_{z} \times \sqrt{\tau}}{\sqrt{3}}$ & $1/2$ & $ \sigma\left(3 \right)$\\ 
\hline
\rule[-1ex]{0pt}{3.5ex}  Drift rate ramp &  $\frac{Q_{z} \times \tau}{\sqrt{2}}$ & 1 & $\sigma\left(\sqrt{2}\right)$ \\
\hline
\end{tabular}
\end{center}
\end{table} 

\subsection{PSD ANALYSIS}
\label{sec:psd}

\begin{figure}[ht]
\centering
\includegraphics[width=3.6in]{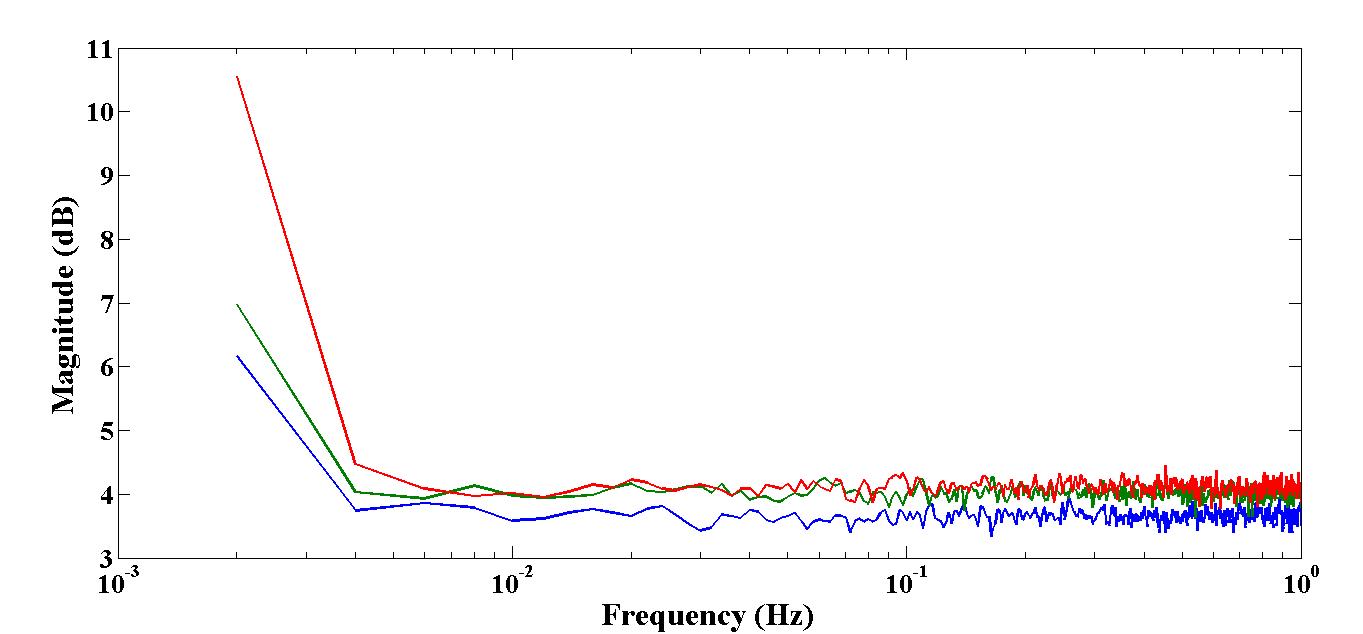}
\includegraphics[width=3.6in]{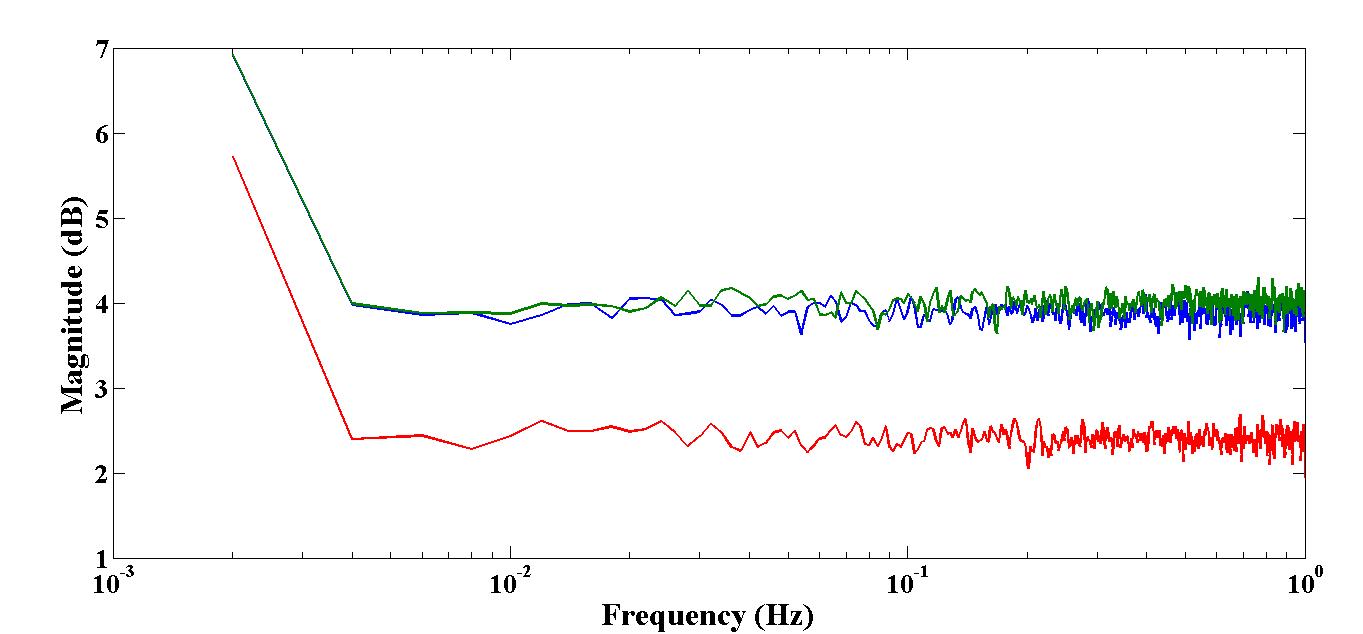}
\caption{Results of the PSD analysis of the accelerometer ({\it Top}) and gyroscope ({\it Bottom}) data. Slopes of these curves were measured by linear fitting.}
\label{fig:psd}
\end{figure}

We have performed power spectral density analysis on the IMU data to determine the colour of the noise. PSD analysis gives the amount of energy present at a particular frequency. A periodogram, a Welch method, and an FFT are the commonly used techniques to estimate the PSD of a signal. We used the Welch [\citenum{Kurt}] method to determine the PSD of the signal. In this method, the Fourier transform of the autocorrelation of a small overlapping section within the signal is taken.
\begin{equation}
S(f) \xrightarrow{\mathcal{F}} R(\tau)\,,
\end{equation}
where $S(f)$ is the PSD of the signal and $ R(\tau)$ is the autocorrelation function. These results are averaged to obtain an estimate of the PSD. We found the PSD of the accelerometer and gyroscope $X$, $Y$ and $Z$ axes data (Fig.~\ref{fig:psd}). The colour of the noise was found from the slope of the PSD (third column in Table~\ref{tab:AV_AN}). The slope of the PSD curve close to zero indicates the white noise, in agreement with the result obtained from Allan Variance analysis (Sec.~\ref{sec:av}).

\subsection{PDF ANALYSIS}
\label{sec:pdf}

We performed the PDF analysis on the stationary IMU data to understand the random nature of the measurement noise. The accelerometer and gyroscope data were collected over the period of one hour. The same experimental setup as in the previous analysis (Fig.~\ref{fig:expt_setup}) was used here. The mean value of the signal was subtracted from the collected data to obtain the noise. We plotted the histogram of the noise signal and tried to fit it by the different distributions. We found that the noise was best fit by a Gaussian distribution. A Gaussian distribution is completely represented by its mean and standard deviation. All the fitted Gaussian distributions had zero mean, and the standard deviations are given in  Table~\ref{tab:AV_AN} (last column). From the AV and PDF analysis we found that the noise in IMU is white Gaussian in nature.

\begin{table}[h!]
\caption{Results of the IMU data analysis by different methods (Allan variance, PSD and PDF)} 
\label{tab:AV_AN}
\centering
\begin{tabular}{|l|c|c|l|}
\hline
\rule[-1ex]{0pt}{3.5ex}  MPU-9050 & ADF slope & PSD slope & std \\
\hline
\rule[-1ex]{0pt}{3.5ex}  Gyroscope $X$ axis & $-0.47\,\, (^{\circ}$/s$^{2}$)/s & $-0.02$& $0.67\, ^{\circ}$/s \\
\hline
\rule[-1ex]{0pt}{3.5ex}  Gyroscope $Y$ axis & $-0.49 \,\,(^{\circ}$/s$^{2}$)/s&$0.02$ & $0.78\,^{\circ}$/s\\
\hline
\rule[-1ex]{0pt}{3.5ex}  Gyroscope $Z$ axis & $-0.47\,\, (^{\circ}$/s$^{2}$)/s& $-0.03$ & $0.12\,^{\circ}$/s\\
\hline
\rule[-1ex]{0pt}{3.5ex} Accelerometer $X$ axis & $-0.53$ (m/s$^{2}$)/s& $0.01$& $0.04\,$m/s$^2$ \\
\hline
\rule[-1ex]{0pt}{3.5ex} Accelerometer $Y$  axis &   $-0.48$  (m/s$^{2}$)/s& $-0.01$& $0.06\,$ m/s$^2$ \\
\hline
\rule[-1ex]{0pt}{3.5ex} Accelerometer $Z$ axis &  $-0.54$ (m/s$^{2}$)/s & $-0.01$ & $0.07\,$m/s$^2$\\
\hline
\end{tabular}
\end{table}

\section{FILTERING THE ERRORS}
\label{sec:filter}

We implemented different low-pass filters in Arduino controller to remove the noise from the IMU response and to obtain accurate version of the attitude data.

\subsection{MOVING AVERAGE FILTERS}

\begin{figure}[ht]
\centering
\includegraphics[scale=0.25]{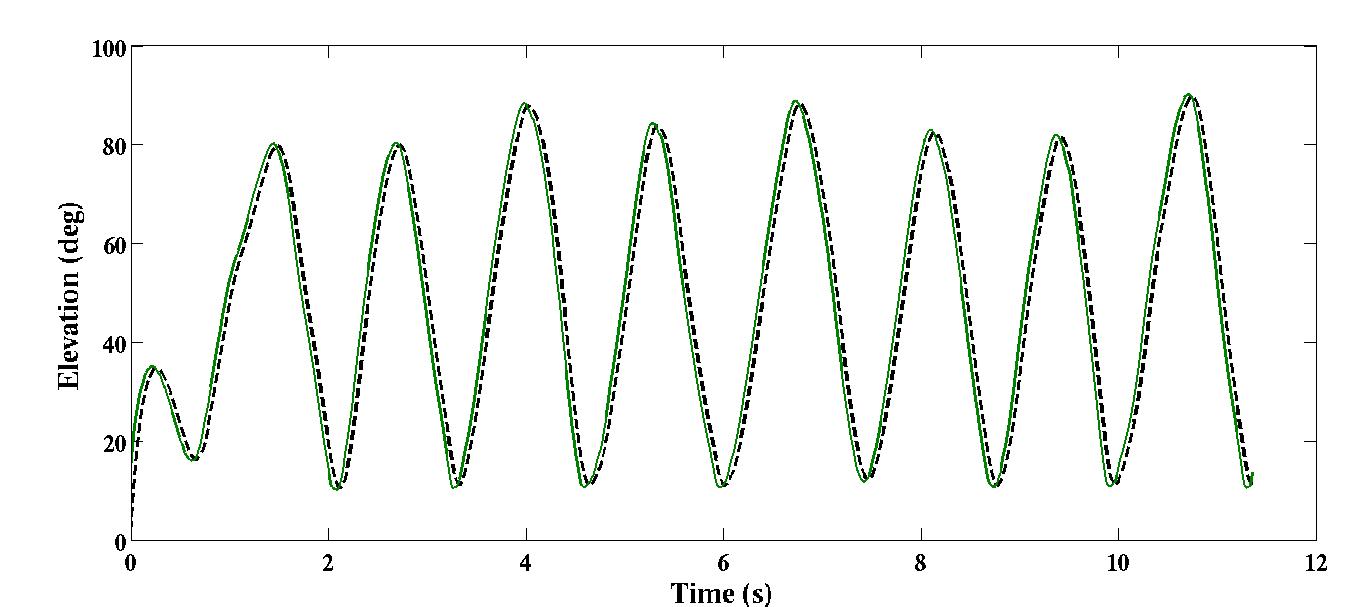}
\includegraphics[scale=0.25]{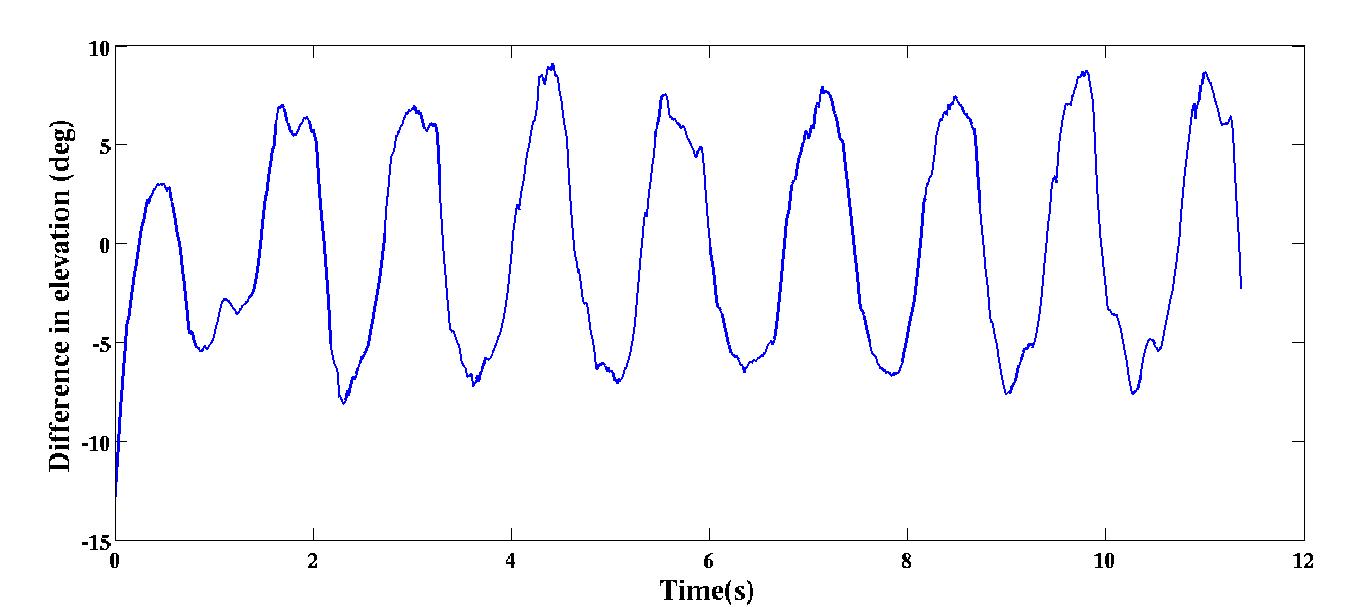}
\caption{{\it Top}: Performance of the moving average filter. green continuous line is the unfiltered data and black dash line is the filtered response. {\it Bottom}:  The offset between the filtered and unfiltered data.}
\label{fig:moving_avg}
\end{figure}

Moving average filter [\citenum{Grace}] was implemented to filter the IMU signal in real time in the Arduino controller. Moving average filter takes a fixed subset of the data points from a given a series of data, and the first element of the moving average is obtained by taking the mean of the fixed subset. The second element of moving average is found by excluding the first number of the subset, taking the next number following the original subset in the series and finding its mean. This process is repeated over the entire data series to obtain the moving averaged version of the data. We implemented a moving average filter of subset size of 10 samples in our Arduino controller to reduce the time lag between the input and output signal from the filter. Even though this filter gave a reasonably good smoothing response, the response of the filter was continuously lagging behind the input (Fig.~\ref{fig:moving_avg}, {\it Top}) with variable offsets. We found the maximum offset between the filtered and unfiltered data to be approximately $9^{\circ}$ (Fig.~\ref{fig:moving_avg}, {\it Bottom}). Therefore, we discarded this filter and tried another low-pass filter -- Savitzky-Golay (SG) filter -- for real-time filtering of the IMU data.

\subsection{SAVITZKY-GOLAY FILTER}

\begin{figure}[ht]
\centering
\includegraphics[width = 0.35\textwidth,height = 4cm]{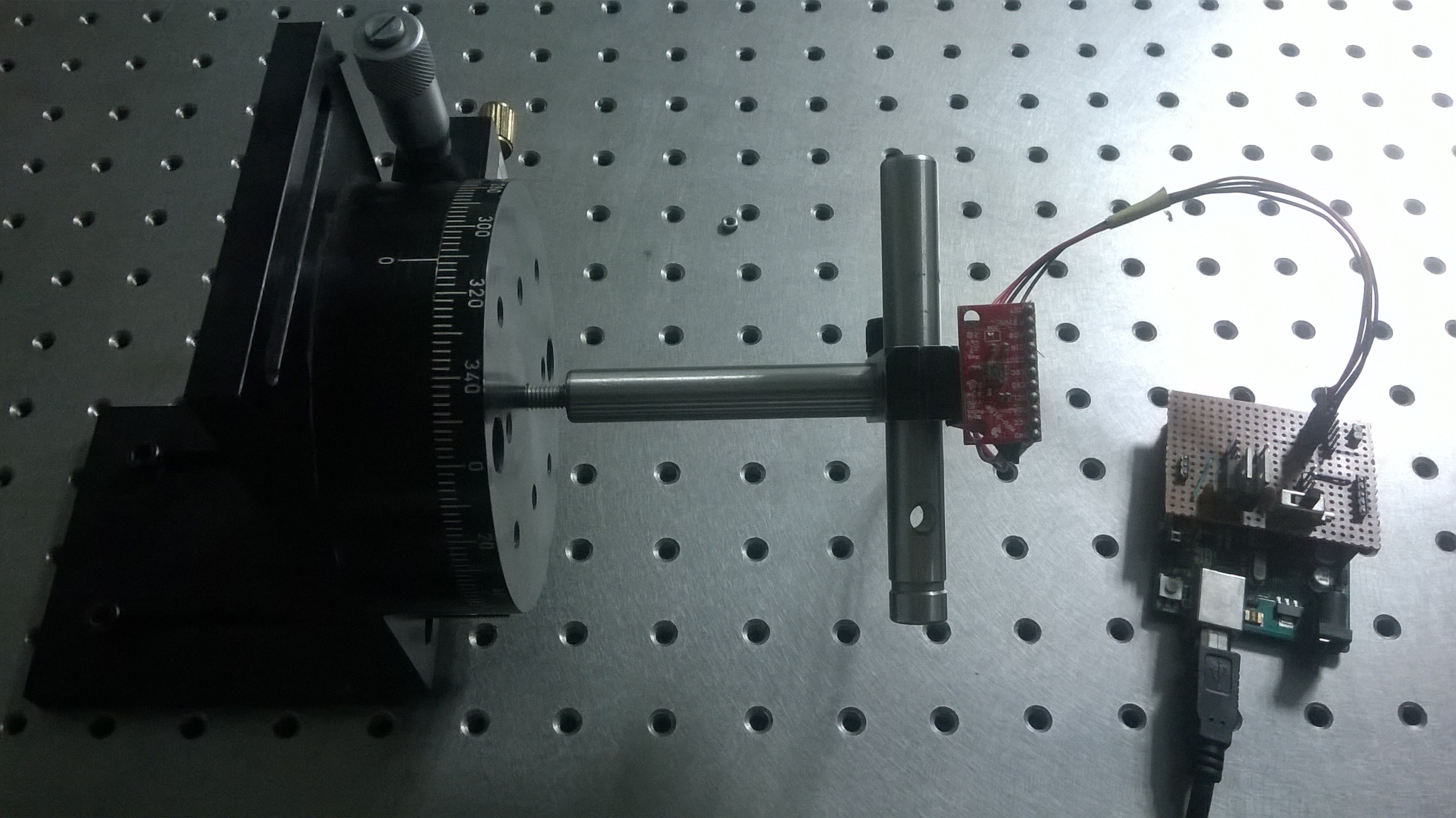}
\caption{Experimental setup used to measure the performance of the SG filter.}
\label{fig:imu_cal_set}
\end{figure}

\begin{figure}[ht]
\centering
\includegraphics[scale=0.3]{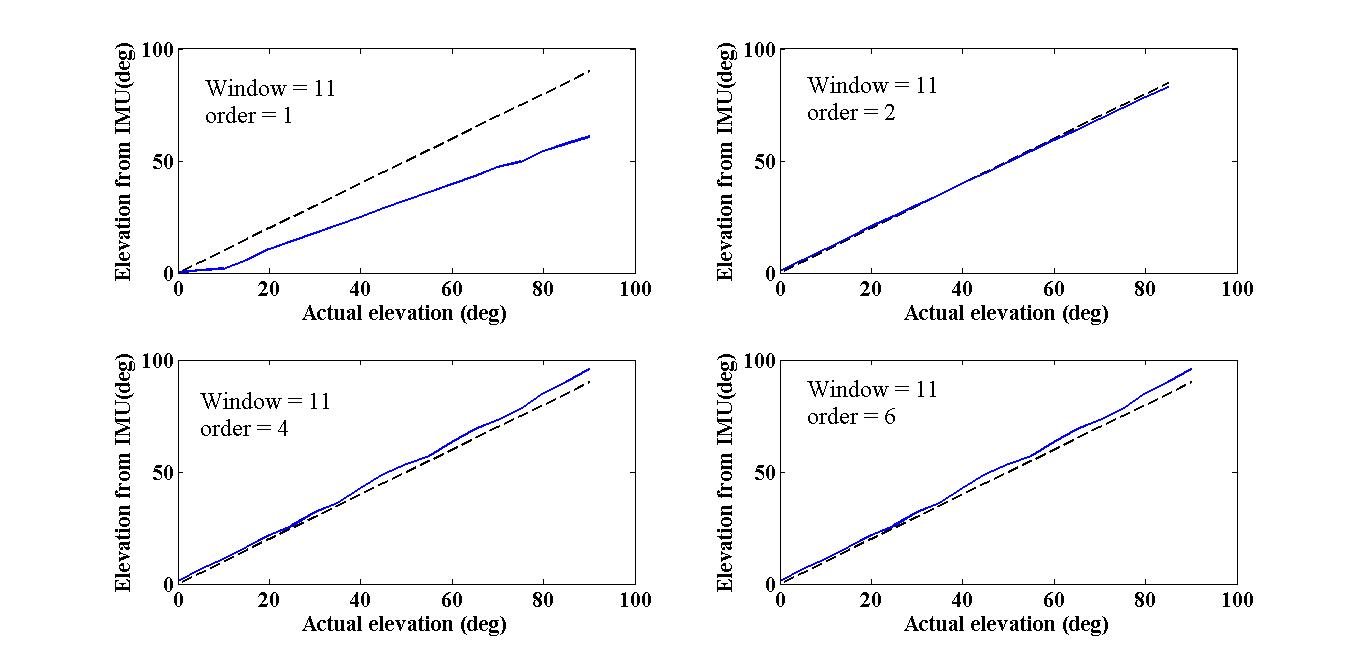}
\caption{Responses of different SG filters. The window size and polynomial order fitted for the data point are displayed inside each plot. The broken line is the theoretical curve, and the solid line is the response obtained from SG filter. In most cases, the  response of the SG filter have considerable deviation from the actual value.}
\label{fig:sg_calib}
\end{figure}

Another method to smooth the data is by using the SG filter [\citenum{Schafer}]. Here, the filtering is done by fitting the $n$-th order polynomial $P^{n}(x)$ to a sample of $2M+1$ data points. In the  SG filter algorithm, these 2M+1 samples are selected in such a way that they are indexed from $-M$ to $+M$. The first element in the SG filtered data is $P^{n}(0)$. The next element is found by modifying the initial data set by a shift,  forwarding by $M$ elements and repeating the same process. We mounted our IMU on the $0.5^{\circ}$ resolution rotational stage and moved it in elevation (see Fig.~\ref{fig:imu_cal_set} for the experimental setup). Only the elevation readings were collected by the Arduino controller. We implemented eleven-element window SG filter in the Arduino to filter the IMU noise. The level of smoothing of the data can be changed by changing the window size and the order of the polynomial. We tried different polynomial orders, i.e. $1,2,4,6$, to find a suitable filter; SG-filtered version of elevation data is shown in  Fig.~\ref{fig:sg_calib}. We found that in real-time applications, it is not possible to filter the signal using the SG filter, because the filter data and the sensor data have considerable deviation.

\subsection{KALMAN FILTER}

\begin{figure}[ht]
\centering
\includegraphics[scale=0.3]{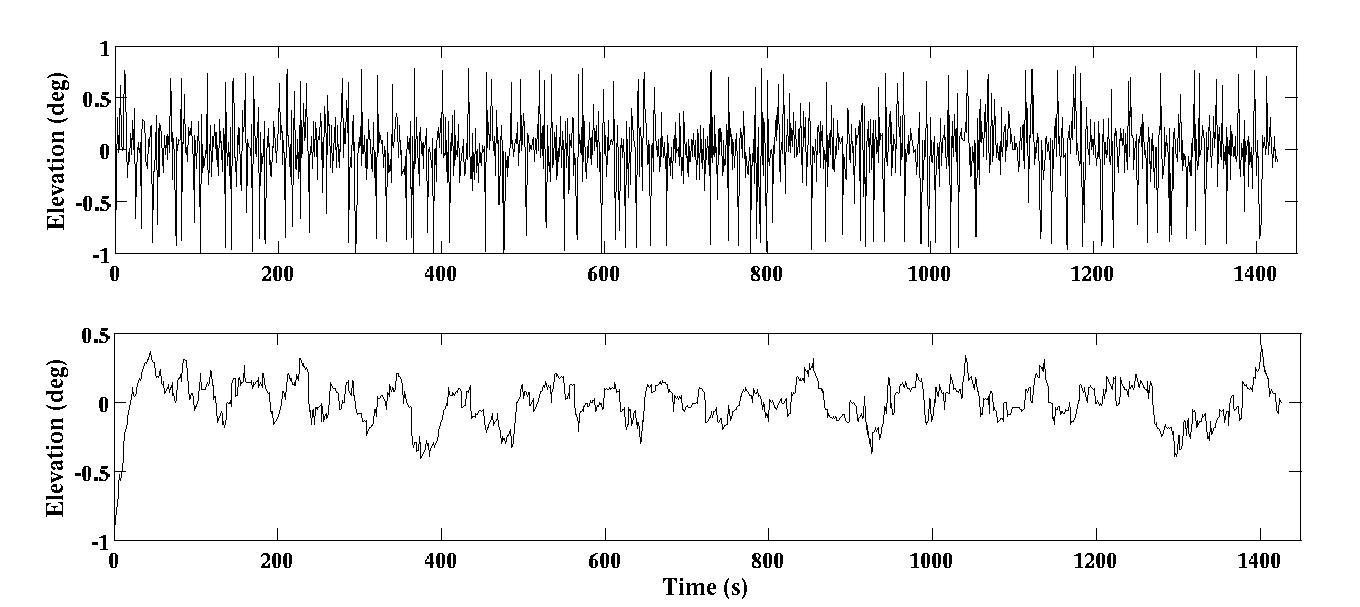}
\caption{{\it Top}: Elevation calculated from the accelerometer versus time. {\it Bottom}: Sensor-fused elevation data from accelerometer and gyroscope after the Kalman filter versus time.}
\label{fig:sensfusion}
\end{figure}

A Kalman filter [\citenum{Lefferts}] is implemented in the controller to remove the noise form the IMU since our smooth filter didn't give satisfactory results. The gyroscope gives precise values over short time duration but drifts for longer observations [\citenum{Sreejith14}], while the converse is true for the accelerometer: there is no drift over long periods of time but a significant jitter occurs on short time scales ($0.330^{\circ}$ in 100 ms bins). We can reduce this jitter by combining the data from the gyroscope and the accelerometer using a Kalman filter\footnote{{\tt https://github.com/TKJElectronics/}.}. This is shown in Fig.~\ref{fig:sensfusion}, where the elevation, calculated from only the accelerometer data, is plotted in the top panel, and the elevation from the fusion of the two sensors is plotted in the bottom panel. The deviation in the elevation is reduced to $0.143^{\circ}$ per 100 ms. therefore, we conclude that Kalman filter gave satisfactory results in removing the high-frequency jitter in the IMU response. 

\section{CALIBRATION OF THE IMU AND CONCLUSION}
\label{sec:calib}
\begin{figure}[h!]
\centering
\includegraphics[width = 0.7\textwidth,height = 6cm]{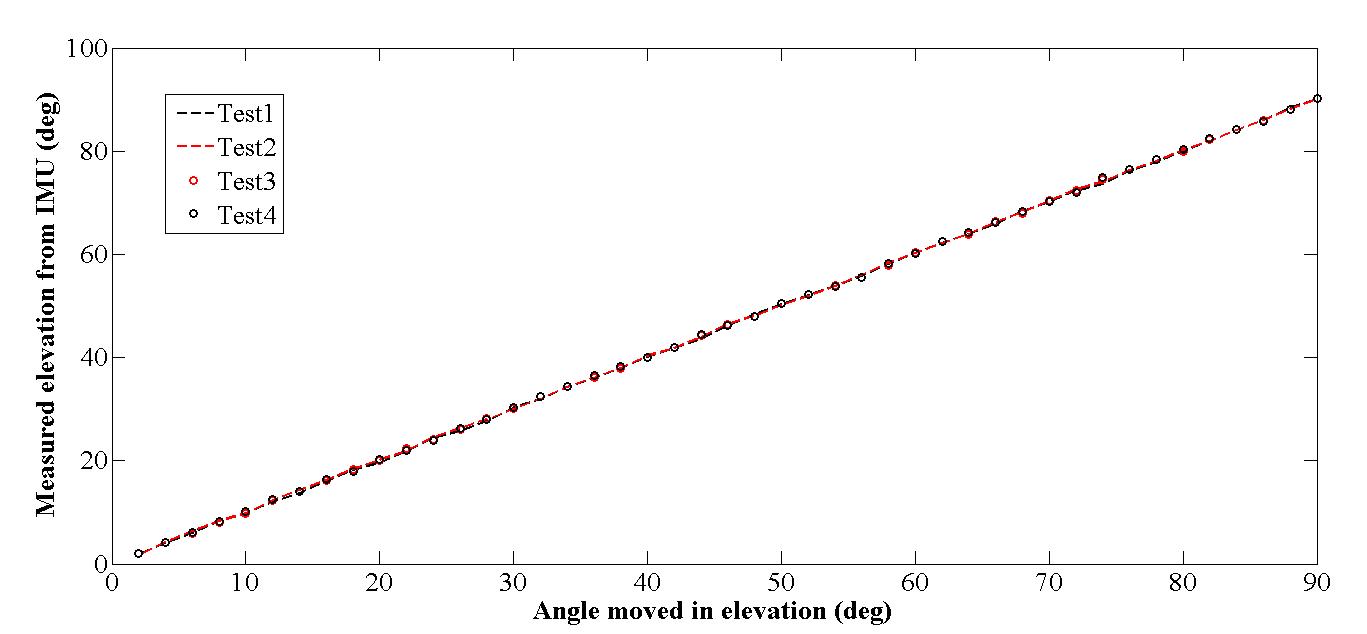}
\caption{Results of the IMU calibration.}
\label{fig:imu_cal_result}
\end{figure}

We calibrated the Kalman-filtered IMU response using the same rotational stage as shown in Fig.~\ref{fig:imu_cal_set}. 
IMU was tilted to a known angle in elevation (actual elevation) using an optical rotational stage having accuracy $0.5^{\circ}$, and IMU reading were collected by the Arduino controller. The elevation values were calculated from the IMU response by Eqs.~\ref{eq:omega} and \ref{eq:alt}. This test was carried out several times (Fig.~\ref{fig:imu_cal_result}), and the average RMS deviation of measured elevation from actual was found to be $0.26^{\circ}$.

In summary, we analyzed IMU measurement noise using the AV and and PSD techniques and found it to be white Gaussian in nature. We characterized it deeper by the PDF analysis. A moving average and a Savitzky-Golay filters were implemented in the controller to reduce the noise, but we found these filters not efficient enough for our purpose. We, however, obtained satisfactory results using a Kalman filter along with the sensor-fusion algorithm, and this will be used in the flight model.

\bibliographystyle{spiejour} 
\bibliography{report} 

\listoffigures
\listoftables
\end{document}